# Organization of Multi-Agent Systems: An Overview

Hosny Ahmed Abbas[1, *], Samir Ibrahim Shaheen[2], Mohammed Hussein Amin[1]

[1]Department of Electrical Engineering, Assiut University, Assiut, Egypt
[2]Department of Computer Engineering, Cairo University, Giza, Egypt

**Email Address:**
hosnyabbas@aun.edu.eg (H. A. Abbas), sshaheen@eng.cu.edu.eg (S. I. Shaheen), mhamin@aun.edu.eg (M. H. Amin)

**Abstract:** In complex, open, and heterogeneous environments, agents must be able to reorganize towards the most appropriate organizations to adapt unpredictable environment changes within Multi-Agent Systems (MAS). Types of reorganization can be seen from two different levels. The individual agents level (micro-level) in which an agent changes its behaviors and interactions with other agents to adapt its local environment. And the organizational level (macro-level) in which the whole system changes it structure by adding or removing agents. This chapter is dedicated to overview different aspects of what is called MAS Organization including its motivations, paradigms, models, and techniques adopted for statically or dynamically organizing agents in MAS.

## 1. Introduction

Complexity and highly distribution are the key characteristics of modern real world systems. The complexity of the near future and even present applications can be characterized as a combination of aspects such as great number of components taking part in the applications, knowledge and control have to be distributed, the presence of non-linear processes in the system, the fact that the system is more and more often open, its environment dynamic and the interactions unpredictable [3]. Further, the increasing complexity, heterogeneity, and openness of modern software systems have reached a point that imposes new demands on their engineering technologies. It is expected that conventional engineering approaches will stand powerless in front of future systems increase in scale and complexity either vertically (control and information layers) or horizontally (physical distribution). It doesn't mean that conventional engineering techniques will become obsolete and have to be thrown away. Absolutely, they only need to be integrated with new engineering styles where concepts such as, decomposition, autonomy, modularity, and adaptivity can be collectively combined in one system. MAS are considered as a promising engineering (i.e., architectural) style for developing adaptive software systems able to handle the continuous increase in their complexity as a result of their open, heterogeneous, and continuous evolution nature. They model the system as distributed autonomous agents cooperate together to achieve system goals. The ability of agents to dynamically reorganize to adapt working environment dynamic changes is a key feature provided by MAS. It is obvious that the natural way to model a complex system is in terms of multiple autonomous components that can act and interact in flexible ways in order to achieve their objectives, and also that agents provide a suitable abstraction for modeling systems consisting of many subsystems, components and their relationships [22]. Ferber [23] described how agents, as a form of distributed artificial intelligence, are suitable for use in application domains which are widely distributed. MAS are currently considered as the most representatives among artificial systems dealing with complexity and highly distribution [24]. MAS allow the design and implementation of software systems using the same ideas and concepts that are the very founding of human societies and habits. These systems often rely on the delegation of goals and tasks among autonomous software agents, which can interact and collaborate with others to achieve common goals [34]. In other words, an agent falls somewhere between a simple event-triggered program and one with human collaborative abilities [36].

In contrast to initial MAS research, which concerned individual agents' aspects such as agents' architectures, agents'

mental capabilities, behaviors, etc, the current research trend of MAS is actively interested in the adaptivity, environment, openness and the dynamics of these systems. Also, there is a great attention towards the MAS technique as a way to design self-organized systems. In open environments, agents must be able to adapt towards the most appropriate organizations according to the environment conditions and their unpredictable changes. Agent organizations are considered as an emergent area of MAS research that relies on the notion of openness and heterogeneity of MAS and imposes new demands on traditional MAS models [44]. MAS that have the ability to dynamically reorganize (regardless of the type of reorganization, self or enforced) will be adaptive enough to survive against their dynamic and continuously changing working environments. Dynamic reorganization can take many forms, for instance, agents can dynamically change their roles, behaviors, locations, acquaintances, or the whole system organization structure can be dynamically changed.

An agent organization can also be defined as a social entity composed of a specific number of members (agents) that accomplish several distinct tasks or functions and that are structured following some specific topology and communication interrelationships in order to achieve the main aim of the organization. Thus, agent organizations assume the existence of global common goals, outside the objectives of any individual agent, and they exist independently of agents [64][65].

This chapter is dedicated to provide a comprehensive overview of MAS organization including its motivations, paradigms, and familiar organizational models. The remaining of this chapter is organized as follows: Section 2 explores MAS literature to identify the motivations towards agent organizations. Section 3 presents different approaches and paradigms used to organize agents within multi-agent systems. Section 4 introduces what is called organizational models, which concern the abstractions, languages, approaches and techniques for modeling dynamically reorganized MAS. And Section 5 concludes the article and highlights future work.

## 2. Motivations to MAS Organization

This section is dedicated to identify from MAS literature the suggested motivations to give increasing attention to MAS organization. Basically, a MAS is formed by the collection of autonomous agents situated in a certain environment, respond to their environment dynamic changes, interact with other agents, and persist to achieve their own goals or the global system goals. There are two viewpoints of MAS engineering, the first one is the agent-centered MAS (ACMAS) in which the focus is given to individual agents. With this viewpoint, the designer concerns the local behaviors of agents and also their interactions without concerning the global structure of the system. The global required function of the system is supposed to emerge as a result of the lower level individual agents interactions in a bottom-up way.

Picard et al. [13] stated that the agent-centered approach takes the agents as the "engine" for the system organization, and agent organizations implicitly exist as observable emergent phenomena, which states a unified bottom-up and objective global view of the pattern of cooperation between agents. Further, Picard gives the ant colony [15] as an example, where there is no organizational behavior and constraints are explicitly and directly defined inside the ants. The main idea is that the organization is the result of the collective emergent behavior due to how agents act their individual behaviors and interact in a common shared and dynamic environment.

The key problems of the ACMAS viewpoint are unpredictability and uncertainty. Because the whole is more than the sum of its parts [14], this approach can lead to undesirable emergent behaviors that may impact system performance, as a result, this approach might be not suitable to design and engineer complex multi-agent systems. The MAS applications engineered by the ACMAS approach are closed for agents that are not able to use the same type of coordination and behavior, and that all global characteristics and requirements are implemented in the individual agents and not outside them [10].

Weyns [11] stated that giving the responsibility of system organization implicitly to individual agents, as in the ACMAS approach, in addition to their functional responsibilities is not adequate because it is a type of dual responsibility, which is very complex to engineer and not suitable for handling real world complexity and other emerged characteristics such as highly distribution, unpredictability, uncertainty, and continuous evolution.

The second viewpoint of MAS engineering is what is called organization-centered MAS (OCMAS) in which the structure of the system is given a bigger attention through the explicit abstraction of agent organization. With that approach, the designer designs the entire organization and coordination patterns on the one hand, and the agents' local behaviors on the other hand. It is considered as a top-down approach because the organization abstraction imposes some rules or norms used by agents to coordinate their local behaviors and interactions with other agents.

The OCMAS viewpoint has been promoted by many pioneers in MAS research. For instance, Jennings and Wooldridge [2] stated that MAS contribute to the software engineering (SE) discipline as a way to simplify the design of complex software systems but considering MAS with no real structure isn't suitable for handling current software systems complexity, and higher order abstractions should be used and some way of structuring the society is typically needed to reduce system complexity, to increase system efficiency, and to more accurately model the problem being tackled. Odell et al. [4] stated that the current practice of MAS design tends to be limited to individual agents and small face-to-face groups of agents that operate as closed systems which is not adequate to model and design of complex adaptive systems. Also Gutknecht and Ferber [66] argued that taking organizational concepts, such as groups, roles, structures, dependencies, etc, as first class citizens, and relating them to the behavior of agents is a key issue for building large scale and complex systems. In another article, Ferber [6] also stated that representing a MAS as an organization consists of roles enacted by agents arranged (statically or dynamically) to form groups of agents, can handle many drawbacks such as system complexity, uncertainty, and system dynamism.

Gasser [3] stated that we simply have hardly any real experience building truly heterogeneous realistically coordinated multi-agent systems that work together and almost no basis for systematic reflection and analysis of that experience. Further, Horling et al. [5] stated that our real world getting more complex and highly distributed and that should be reflected in new software engineering paradigms such as MAS. Therefore, the adoption of higher order abstract concepts like organizations, societies, communities, and groups of agents can reduce systems complexity, increase its efficiency, and improve system scalability.

Establishing an organizational structure that specifies how agents in a system should work together helps the achievement of effective coordination in MAS [39]. Broek [7] stated that complexity of real world applications needs to be tackled from higher abstraction order such as organizations which can be used to limit the scope of interactions, provide strength in numbers, reduce or manage uncertainty, and formalize high-level goals which no single agent may be aware of. Further, Hübner [8] confirmed that organizations provide a framework for structuring and managing agents' interactions and serve as a kind of tuning of the agents autonomy level. Furthermore, Burns et al. [12] stated that in organization theory [25][26], it is commonly accepted that different types of organizational structure are suitable for particular environmental conditions and one of the main reasons for creating organizations is to provide stable means for coordination that enable the achievement of global goals.

Moreover, Corkill et al. [36] stated that as agent-based systems become more widespread and complex, designed organization will become an important aspect of effective system performance, and they suggested the possible situations where organization design will be very important such as, large number of agents, long duration of agent activities, more repetitive activities, more activities require shared resources, more collaborative the activities, more specialized agents, less capable agents, and less slack resources are available. Also, they emphasized that no one organization is right for every situation.

In nutshell, proposing a way for statically or dynamically organizing MAS, has been given great attention by MAS researchers, as a promising approach for handling the challenging issue of engineering complex and large-scale software systems. The adoption of the ACMAS or OCMAS viewpoints mainly depends on the nature of application domain and the degree of system complexity. The developers interested in bottom-up self-organized systems will prefer the ACMAS approach and the developers interested in top-down system reconfiguration will prefer the OCMAS approach. In the MAS literature there are two communities each adopts and concerns one of the two engineering approaches. The first one is SASO (Self-Adaptive and Self-Organizing systems) which concerns the ACMAS viewpoint. And the second one is COIN (Coordination, Organization, Institutions and Norms in agent systems) which concerns the OCMAS viewpoint.

The OCMAS viewpoint is more adequate for engineering complex adaptive multi-agent systems, which are expected to be, in the near future, the mainstream approach for engineering large-scale and even ultra-large scale application domains especially with the evolving topic of the Internet of Things (IoT) [17], which concerns devices capable to communicate via the Internet and manipulate an enormous amount of data. Examples of such application domains are CPS (Cyber-Physical Systems) [16], Smart Grids [18], global SCADA (Supervisor Control and Data Acquisition) [19], Pervasive Computing [20], Ubiquitous Computing [21], etc. The next section explores the familiar paradigms of MAS organization.

## 3. Paradigms of MAS Organization

Originally, the organization abstract is inspired from business human organizations, which are constituted of a number of roles, so a key concept in the design of OCMAS is that of roles, which define normative behavioral repertoires for agents [4]. A role is defined as an abstract description of some activity or functionality, for instance in a business human organization we may see a role like Manager who is responsible of the organization management and the coordination between other organization members (roles). In MAS, agents are supposed to enact roles according to the capabilities of each agent. It is also possible that one agent can enact many roles in the same time. The role enacted by an agent has a direct effect on the agent behavior and interaction with other roles (agents). Odell et al. [27] described two familiar ways for assigning roles to agents, endogenously by emergent self-organization as the system runs, or exogenously by the system designer when the system is constructed or modified. The adoption of human organization theory was the focus of distributed systems in general before multi-agent systems, which are themselves distributed systems [30][31][32][33].

Modern organizations (real or virtual) are characterized by their complex structure, dense information flows, and incorporation of information technology, they also characterized by highly dynamic, constantly changing, organic structure and show hardly identified, not formalized, non-linear behavior [28][29]. These challenges enforce the urgent need to a new way of engineering multi-agent systems.

Inspired from human organizations, Galbraith [37] described an agent organization as an entity that is composed of a set of agents, working together to achieve a shared purpose through a division of labor, integrated by decision processes continuously through time. Further, Galbraith pointed out that an organization consists of patterns of behavior and interaction that are relatively stable and change slowly over time.

Shehory [1] defined MAS organization as the way in which multiple agents are organized to form a multi-agent system. The relationships and interactions among the agents and specific roles of agents within the organization are the focus of multi-agent organization. The use of organizations provides a new way for describing the structures and the interactions that take place in MAS. Dignum [10] stated that agent organization can be understood from two perspectives: organization as a process and organization as an entity. In other words, organizations can be considered as the process of organizing a set of individual agents, thus in this sense it is used to refer to constraints (structures, norms and patterns) found in a social context that shape the actions and interactions of agents [53]. In other situations, it can be considered as an entity in itself, with its own requirements and objectives and is represented by (but not identical to) a group of agents. In fact, agent organizations demand the integration of both perspectives and rely for a great extent on the notion of openness and heterogeneity of MAS.

Figure 1 illustrates how a MAS can be seen from two levels, the individual agents' level and the organizational level. The organizational level presents a higher order abstraction of the lower agents' level.

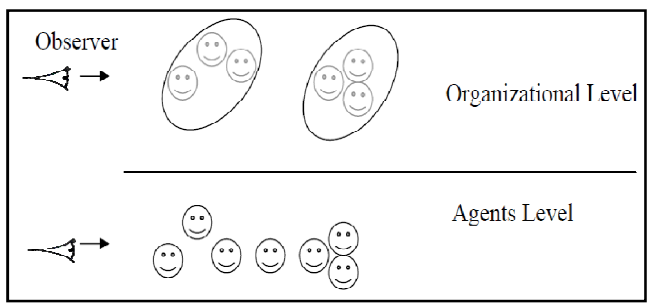

*Figure 1. Organizational level vs. individual level in MAS.*

Ferber et al. [6] proposed a set of general principles that should be taken into account when designing MAS with organizational dimension:

1. The organizational level describes the "what" and not the "how". In other words, the organizational level imposes a structure into the pattern of agents' activities, but does not describe how agents behave.
2. No agent description and therefore no mental issues at the organizational level. The organizational level should not say anything about the way agents would interpret this level.
3. An organization provides a way for partitioning a system, each partition (or agent group) constitutes a context of interaction for agents. Thus, a group is an organizational unit in which all members are able to interact freely.

Ferber principles provide important general guidelines for OCMAS research. They identify precisely the logical relation between agents and their organization regardless of the nature of organization (i.e. a process or an entity). The first principle concerns the autonomy of agents. Agents should be autonomous but they may be guided by some general organizational norms or constraints. Full autonomy is not a preferred agent characteristic in MAS research, we can only find a type of full autonomy with humans because they have perfect rational minds, but agents (software or hardware) designed for specific missions in certain application domains and the concept of safety imposes some constraints on agents' autonomy, in these situations, a designed organization, where agents give up some degree of self-motivation and autonomy can be an appropriate choice [36].

The second principle concerns the unawareness of agents about the existence of the organizational level, which according to Ferber should be transparent from agents. In other words, agents should be affected indirectly by the change of system organization (i.e., through environment). The third principle concerns system modularity. Organizations provide a way for, statically or dynamically, decomposing the system. Modularity and flexibility of system decomposition enhance system maintainability.

Horling and Lesser [5] also stated that organizational design employed by an agent system can have a significant, quantitative effect on its performance characteristics, and they surveyed the major organizational paradigms used in multi-agent systems. These include hierarchies, holarchies, coalitions, teams, congregations, societies, federations, markets, and matrix organizations. Also, they provided a description of each paradigm, and discuss its advantages and disadvantages, further, they provided examples of how each organization paradigm may be instantiated and maintained. Table 1 provides a summary of Horling and Lesser [5] work. The Table contains a number of methods by which MAS could be organized and highlights the key characteristics, benefits, and drawbacks of each organization paradigm. Similar work was provided by Carley and Gasser [35]. The main conclusion of these surveys is that no single organization paradigm is necessarily better than all others in all situations. The selection made by a designer should be dictated by the needs imposed by the system's goals, the resources at hand, and the environment in which the participants will exist. In other words, an organization paradigm that can be described as a fit-to-all paradigm does not exist (at least till now!). A MAS can be statically (in design time) organized using any of the organization paradigms presented in Table 1, not only this but also hybrids of these and others in addition to dynamic changes from one organization style to another are also possible [1] with the price of implementation complexity. The later case is called dynamic reorganization which is currently a very active research area within MAS discipline. The next subsections present in more details the concept of dynamic reorganization and its captivating relevant concepts, self-organization and emergence.

### 3.1. Dynamic Reorganization

Earlier proposed MAS organization mechanisms tackled with organizational aspects at design time, that approach requires some important initial knowledge about the exact purposes and objectives of the system-to-be and every interaction to which it may be confronted in the future have to be known in design time [41]. However, the openness, complexity, and heterogeneity of modern software systems impose new demands and requirements on agent-oriented software engineering (AOSE) [71], which is concerned with the development of feasible, effective, and adaptive MAS. Building adaptive MAS (AMAS) able to handle openness, complexity, and highly distribution of modern real world applications has recently attracted great attention.

*Table 1. Analysis of Some of Possible MAS Organization Paradigms (adopted from [5]).*

| Paradigm | Key Characteristic | Benefits | Drawbacks |
| --- | --- | --- | --- |
|  | Decomposition | Maps to many common domains; handles scale well | Potentially brittle, can lead to bottlenecks or delays |
| Holarchy | Decomposition with autonomy | Exploit autonomy of functional units | Must organize holons, lake of predicable performance |
| Coalition | Dynamic, goal-directed | Exploit strength in number | Short-term benefits may not outweigh organization construction costs |
| Team | Group level cohesion | Address larger grained problems; task-centric | Increased communication |
| Congregation | Long-lived, utility-directed | Facilitates agent discovery | Sets may be overly restrictive |
| Society | Open system | Public services; well defined conventions | Potentially complex, agents may require additional society-related capabilities |
| Federation | Middle-agents | Matchmaking, brokering, translation services, facilitates dynamic agent pool | Intermediaries become bottlenecks |
| Market | Competition through pricing | Good at allocation, increased utility through centralization, increased fairness through bidding | Potential for collusion, malicious behaviour, allocation decision complexity can be high |
| Matrix | Multiple managers | Resource sharing, multiple influenced agents | Potential of conflicts, need for increased agent sophistication |
| Compound | Concurrent organizations | Exploit benefits of several organizational styles | Increased sophistication, drawbacks of several organizational styles |

AMAS designed to be capable to adapt themselves to unforeseen situations in an autonomous manner. They can be realized by enabling the system to dynamically reorganize to adapt its environment changes [42]. Dynamic reorganization is a way to design and develop AMAS. It can be described as the change of MAS structure and behavior as a result of internal (local) or external (supervisory) demand. The external demand can be for example human intervention. The internal demand emerges from the system itself as an autonomous system to adapt environments changes. Generally, dynamic reorganization in MAS takes place as a result of individual agents' interactions. However, in many application domains the environment can stimulate MAS reorganization (e.g., when removing or adding environment resources), the system may reorganize to adapt the change of environment. In other words, reorganization is the answer to change in the environment.

Dignum et al. [40] identified two types of MAS dynamic reorganization, emergent Organization in which global behavior cannot be specified in advance, but emerges from the interaction of local behaviors. In other words, agents' interactions may eventually create dynamic organizations [44]. Thus, emergent organizational behavior is primarily a bottom-up process in which agents look for interaction and local control decisions that have been effective in the past and give similar decisions preference in the future. The ACMAS viewpoint concerns this type of reorganization. The other type of reorganization is called designed organization, which has an explicit interaction structure that determines the coordination of the agents participating. Designed systems are created using organization design knowledge and task-environment information to develop an explicit organizational structure, that is then elaborated by the individual agents into appropriate behaviors. Designed organization exhibits predicable and controllable behavior, dynamic change implies the need for highly intelligent and communicative agents (at least some of them) that can reason about and negotiate change. Designed organization is the main concern of the OCMAS approach. In human organizations, it has been proven that designed organizations perform better than those that emerge naturally. This viewpoint holds for agent organizations as well, that is because the global behavior of emergent organizations cannot be predicted and changes cannot be guided, which makes this type less suitable for situations where coordinated and goal-directed global action is required.

Picard et al. [13] added the agents' awareness /unawareness of the existence of the organization structure as a dimension of the organization modification process and he identified four cases:

1. The agents don't represent the organization, although the observer can see an emergent organization. In some sense, they are unaware that they are part of an organization.
2. Each agent has an internal and local representation of cooperation patterns which it follows when deciding what to do. This local representation is obtained either by perception, communication or explicit reasoning.
3. The organization exists as a specified and formalized schema, made by a designer but agents don't know anything about it and even do not reason about it. They simply comply with it as if the organizational constraints were hard-coded inside them.
4. Agents have an explicit representation of the organization which has been defined. The agents are able to reason about it and to use it in order to initiate cooperation with other agents in the system. The agents are able to reason about it and to use it in order to initiate cooperation with other agents in the system.

Case 1 and 2 considered as ACMAS and case 3 and 4 considered as OCMAS. The importance of Picard classification of MAS dynamic reorganization is that nearly most of known reorganization methods fit to a specific case or multiple cases. Similar classification proposed by Sichman et al. [44], but he used the concept of observer in the same position as the agent awareness of the organization. Table 2 provides the global picture of possible types of MAS organization with examples. As shown in the table MAS organization is classified according to the awareness/unawareness of individual agents about the presence of the organizational level.

*Table 2. The global picture of MAS organization.*

|  | **Agents unaware** | **Agents aware** |
|---|---|---|
| ACMAS=Emergent organization | Organization is observed. It is implicitly programmed in agents, interactions, and environment. | Organization is observed. Coalition mechanisms programmed in the agents. |
| Concerned Community | SASO | COIN |
| Examples | Swarm-based systems [72] | Contract-Net Interaction Protocol [73] |
| OCMAS = Designed organization | Organization is a design model. It may be hard coded in the agents. | Organization is programmed in the agents and/or in specialized middleware services. |
| Concerned Community | COIN | COIN |
| Examples | AOSE methodologies such as: MASE [74], INGENIAS [75]. | Organizational models such as: AGR [58] MOISE+ [67] |

Picard also, after finishing his valuable study proposed a comprehensive definition of dynamic reorganization as follows:

"Reorganization is a process, endogenous or exogenous, concerning systems in which organization is explicitly manipulated through specifications, constraints or other means, in order to ensure an adequate global behavior, when the organization is not adapted. Agents being aware of the organization state and structure, they are capable of manipulating primitives to modify their social environment. This process can be both initiated by an external entity or by agents themselves, by reasoning directly on the organization (roles, organizational specification) and the cooperation patterns (dependencies, commitments, powers)."

This definition assumes that the agents are aware of the existence of the organizational level, thus it concerns the OCMAS viewpoint. But, what if agents are unaware of the organization level? According to Picard, in this case the dynamic reorganization process is called self-organization which defined by Picard as follows:

"Self-organization is an endogenous and bottom-up process concerning systems in which only local information and representations are manipulated by agents unaware of the organization as a whole, in order to adapt the system to the environmental pressure by modifying indirectly the organization, therefore by changing directly the system configuration (topology, neighborhoods, influences, differentiation), or the environment of the system, by local interactions and propagation, by avoiding predefined model biases."

This definition states that self-organization represents the ACMAS viewpoint. In a self-organized system, agents are unaware of the organization level, the reorganization process is decentralized, implicit, endogenous, and agents are responsible of the system dynamic reorganization, which is often initiated by an environmental change. In a dynamically reorganizing system where agents are aware of the organization level, this process can be decentralized or not, but always explicit and directly performed by entities (designer or agents) manipulating organizational primitives. Therefore, the awareness is a key dimension added by Picard to identify self-organized MAS. The next section provides detailed review of the self-organization concept in MAS.

### 3.2. Self-Organization

Self-* properties [43] (i.e., self-organization, self-healing, self-adaptation, self-configuration, etc) are the most captivating concepts recently appeared in software engineering. They remind us of Einstein ideas about Time Machine, which was and still a far dream of human to travel through time. Human also dreams to design a system, regardless of its nature (software or hardware), able to do all things by itself. A system has all known self-* properties will be amazing. Actually, this type of systems is imaginary (at least till now!); we can only see this system in science fiction movies. However, it is possible to design systems with one or more of self-* properties for predefined purposes and under certain circumstances.

The first use of the term self-organization returns to Ashby [56], in 1947, he stated that a system is said to be self-organized if it changed its own organization rather than being changed by an external entity. Self-organization in software systems received great attention since the last few years. It is an attractive way to handle the dynamic requirements in software in general and MAS in specific. It refers to a process where a system changes its internal organization to adapt to changes in its goals and its working environment without explicit external control. Understanding the mechanisms that can be used to model, assess and engineer self-organizing behavior in MAS is currently an issue of major interest [38].

Picard's definition of self-organization (see previous section) can be rephrased as follows: Self-organization is a process where some form of overall order or coordination arises out of the local interactions between the components of an initially disordered system. This process is spontaneous: it is not necessarily directed or controlled by any agent or subsystem inside or outside of the system. It is often triggered by random fluctuations, which are triggered and amplified by positive feedback. The resulting organization is wholly decentralized or distributed over all the components of the system, it is typically very robust and able to survive and self-repair substantial damage or perturbations.

The roots of the term self-organization return to the work of Glansdorff and Prigogine [46] through thermodynamics studies. They discovered that open systems decrease their entropy (order comes out of disorder) when an external energy is applied on the system. Matter organizes itself under this external pressure to reach a new state where entropy has decreased.

Nature is full of self-organization forms and patterns, for instance social behavior of insects like ants or termites, which formed as a result of indirect communication through environment without the need for any type of direct interaction, this type of interaction is called Stigmergy. Social behavior of humans is also self-organized and gives rise to emergent complex global behaviors. Human beings typically work with local information and through local direct or indirect interactions producing complex societies [45].

Researchers from variety of disciplines who were interested in self-organization in nature found MAS as the adequate engineering style for modeling and simulation of the self-organization phenomena and after a period of time the situation reversed as the MAS researchers, who are concerned with AOSE gave a greet attention to bio-inspired models for developing complex, open, and heterogeneous MAS-based applications. Self-organization and emergence are currently the main focus of AOSE researchers. The adoption of naturally inspired methods and approaches for engineering self-organized MAS is currently a very active research area [47][48]. Mechanisms such as direct interactions [49], Stigmergy [50], reinforcement [51], and agents' cooperation [52] are widely used to design MAS with self-organization behavior.

Another relevant and interesting concept is that of emergence, which can be considered as a process takes place in complex systems (which may or may not be self-organized). Self-organization results from emergence, but there is no guarantee that a self-organized system will always generate emergent phenomena. Understanding how to engineer systems that are capable of presenting self-organized behavior and desirable emergence is currently a very active research area too. The next section introduces briefly the concept of emergence.

### 3.3. Emergence

A lot of confusion exists about the meaning of the two relevant terms emergence and self-organization. One of the sources of the confusion comes from the fact that a combination of both phenomena often occurs in dynamical systems [53]. In MAS domain, self-organization and emergence concepts are recently getting great focus as a way to engineer open, heterogeneous, and complex MAS-based applications such as complex adaptive systems (CAS) [54], which are fluidly changing collections of distributed interacting components that react to both their environments and to one another. The familiar definition of emergence is as a phenomenon where global behavior arises from the interactions between the local parts of the system. This general and vague definition indicates that still there is no consensus of a clear definition for emergence. Also, it indicates the absence of clear understating of its nature. In contrast to the reductionism theory [55], which allows a system to be reduced to the sum of its parts, the emergent global behavior cannot be predicted by observing its parts local behaviors. An accepted operational definition of emergence was proposed by De Wolf and Holvoet [53] as follows:

"A system exhibits emergence when there are coherent emergents at the macro-level that dynamically arise from the interactions between the parts at the micro-level. Such emergents are novel with respect to the individual parts of the system."

This definition uses the concept of an 'emergent' as a general term to denote the result of the process of emergence, i.e., properties, behavior, structure, patterns, etc. The 'level' mentioned refers to certain points of view. The macro-level considers the system as a whole and the micro-level considers the system from the viewpoint of the individual entities that make up the system. The concept of emergence is very complex and it is not fall in the scope of this article, interested readers are invited to explore the emergence relevant references.

### 3.4. Discussion

Static design of MAS is not adequate for modern real world applications, which characterized by their increasing complexity heterogeneity, and openness. Even closed systems in which the number of agents is constant with time should have a type of adaptive dynamic behaviors. All possible behaviors of modern systems cannot be captured at design time and that requires these systems to be adaptive able to adapt changes in their working environments. Dynamic reorganization is currently a familiar way for developing adaptive MAS. As shown in Section 2, the adoption of organizational aspects within MAS is promoted and recommended by pioneers of MAS research. Dynamic reorganization can be described as the change of MAS structure and behavior as a result of internal or external demand. The external demand can be for example human intervention. The internal demand emerges from the system itself as an autonomous system to adapt environments changes. Self-organization is a dynamical and adaptive process where systems acquire and maintain structure themselves, without external control. In self-organizing systems, robustness is used in terms of adaptivity in the presence of perturbations and change. A self-organizing system is expected to cope with that change and to maintain its organization autonomously.

Emergence emphasizes the presence of a novel coherent macro-level emergent (property, behavior, structure, etc) as a result of the interactions between micro-level parts. A combination of emergence and self-organization is a promising approach to engineer large-scale multi-agent systems. In most systems that are considered in MAS literature, emergence and self-organization occur together. Research in MAS and CPS communities focuses on such systems. In very complex (multi-agent) systems, i.e. distributed, open, large, situated in a dynamic context, etc., the combination of emergence and self-organization is recommended.

When a researcher proposes an approach to dynamically reorganize a multi-agent system to adapt environments' changes, he actually proposes what the MAS domain consensus agreed to call as an Organizational (or organization) Model. MAS organizational models will play a critical role in the development of future larger and more complex MAS. The main concern of organizational models is to describe the structural and dynamical aspects of organizations [9]. They have proven to be a useful tool for the analysis and design of multi-agent systems. Furthermore, they provide a framework to manage and engineer agent organizations, dynamic reorganization, self-organization, emergence, and autonomy within multi-agent systems. The next section introduces organizational models in some details.

## 4. Organizational Models

Organizational models have been recently used in agent theory for modeling coordination in open systems and to ensure social order in MAS applications [64]. The adoption of organizational models is currently given great importance within most agent-oriented software engineering methodologies. The motivation to this direction is that in open environments, agents must be able to adapt towards the most appropriate organizations according to the environment conditions and their unpredictable changes. As a result, organizational models should guarantee the ability of organizations to dynamically reorganize as a response to dynamic environment changes. Organizational models are responsible of how efficiently and effectively organizations carry out their tasks, they have been recently used in agent theory for modeling coordination in open systems and to ensure social order in multi-agent system applications [7].

From the business management discipline an organizational model, also called as organizational structure, defines an organization through its framework, including lines of authority, communications, duties and resource allocations. A model is driven by the organization's goals and serves as the context in which processes operate and business is done. The ideal model depends on the nature of the business and the challenges it faces. In turn, the model determines the number of roles needed and their required skill sets. In MAS, the purpose of an organizational model is to enhance the analysis and design of OCMAS, so it's usually integrated with a particular agent-based software engineering methodology.

Before exploring some of the familiar organizational models proposed for modeling complex MAS, it is a suitable time to show the difference between them and MAS development methodologies. In general, a methodology is a body of methods employed by a discipline. A method is a procedure for attaining something. A methodology aims to prescribe all the elements necessary for the development of a software system [57]. AOSE community concerns creating development methodologies suitable for the development of agent-oriented or agent-based software. Typically, a development methodology (agent-oriented or not) comprises an ordered set of phases such planning, analysis, design, implementation, validation, and deployment. An organizational model is a tool adopted within a development methodology for modeling the system-to-be. Typically, it starts in the analysis phase but can expand through design and implementation phases, or in other cases it can expanded through the whole development life cycle.

The next section explores some of familiar proposed MAS organizational models focusing in their tackled organizational aspects, their advantages, and their disadvantages.

### 4.1. Familiar Organizational Models

There is a lot of MAS organizational models proposed in the literature; each of them tackles MAS organization from a different viewpoint. Some of them adopt the ACMAS viewpoint, others adopt the OCMAS viewpoint, and some adopt a hybrid approach concerns both ACMAS and OCMAS viewpoints. In what follows, three of familiar organizational models are introduced.

#### 4.1.1. AGR and AGRE

Ferber et al. [6] proposed a very concise and minimal OCMAS model called AGR, for Agent/Group/Role, also known as the AALAADIN model [58]. The authors of AGR model proposed a set of notations and a methodological framework to help the designer to build MAS using AGR. Further, they presented a set of diagrams (organizational structure, cheeseboard diagram, and organizational sequence diagrams), which may represent the different aspects (static and dynamic) of OCMAS. Their model is based on the dynamic creation of agents groups (agents partitioning) and dynamic forming of hierarchies of groups (Holarchies). They pointed out that their AGR-based model can be integrated with Gaia [59] MAS development methodology to complete the analysis and design phases of MAS development. Figure 2 presents the AGR meta-model.

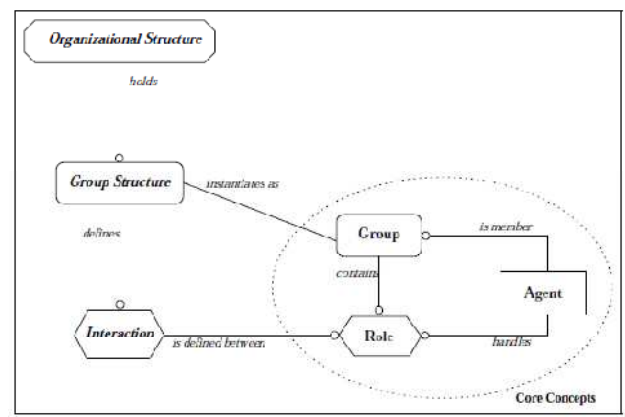

*Figure 2. The AGR Meta-Model.*

The core concepts on which the AGR model is based are agent, group and role. The agent in AGR is assumed to be an active, communicating entity which plays roles within groups, with no restrictions on its internal architecture. The group is defined as the basic unit of agent aggregation. Each agent is part of one or more of these groups. The role is an abstract representation of an agent function, service or identification within a group. Each agent can handle several roles, and each role handled by an agent remains local to a group. Other important abstract concepts are also shown the AGR meta-model shown in Figure 9.1, they are Group Structure and Organizational Structure. The group structure is an abstract representation of the roles required in this group and their interaction relationships and protocols. The organization structure is the set of group structures expressing the design of a multi-agent organizational scheme.

In other paper, Ferber et al. [60] presented an extension of the AGR organizational model, called AGRE (AGR + Environment), which includes physical (or simply geometrical) environments. This extension is based on the concept of a space which can be seen either as a physical area or as a social group.

The main advantages of the AGR/AGRE models are: supporting of heterogeneous agents architectures, heterogeneous communication languages, and dynamic role-group relationships. On the other hand, the disadvantages are multi-role agents which make agents internally complex; roles sharing between Groups can cause overloaded agents, agents ask to join groups which require highly knowledgeable agents, use of mediator agents (brokers) which can be a source of bottlenecks, and very few known real applications.

### 4.1.2. MOISE

Hannoun et al. [67] proposed MOISE (Model of Organisation for multI-agent SystEms); for modeling organizational aspects of MAS. Similar to the AGR model, their model is based on three major concepts: roles, organizational links (roles relations), and groups. What distinguishes the MOISE model is that it tries to integrate both viewpoints, ACMAS and OCMAS. By this way MOISE gives the chance to the designer to model totally or partially the social behavior of system agents by specify the possible organizational structures and that will be useful for system verification and validation. On the other hand, for the sake of flexibility, agents should be able to reason about their social behaviors and have a direct influence on system dynamic reorganization.

The MOISE model is structured along with three levels: (i) for each agent, definition of the tasks that it is responsible of (individual level), (ii) aggregation of agents in large structures (aggregate level), (iii) global structuring and interconnection of the agents and structures with each other (society level). The organization in MOISE is viewed as a normative set of rules that constrains the agents' behaviors [67].

The MOISE organizational model was extended by Hübner et al. to MOISE$^+$ [68] to create an organization-centered model for independently specify the structural and functional aspects then link them by a deontic aspect. Another extension to MOISE$^+$ [69] was done to add dynamic reorganization process to adapt environment changes.

### 4.1.3. MACODO

Weyns et al. [61] presented an organizational model for context-driven dynamic agent organizations. The model defines abstractions that support application developers to describe dynamic reorganization. The organizational model is part of an integrated approach, called MACODO (Middleware Architecture for COntext-driven Dynamic agent Organizations); in this model, the life-cycle management of dynamic organizations is separated from the agents, organizations are first-class citizens, and their dynamics are governed by laws. Moreover, the authors provided a formal specification to describe and specify the semantics of their organizational model abstractions using Z specification language [62], which is based on set theory and first order predicate calculus. The main concern of MACODO is to directly relate organization dynamics to context changes in the environment.

We argue that the main drawback of the MACODO organizational model is the pure dynamically created organization; we argue that an organization should be tackled from the two perspectives, static and dynamic for the sake of long-term system stability. Dynamically creating and vanishing of organizations without keeping an amount of static behavior can impact system stability and prevents it from reaching an equilibrium state. In other words, there should be an amount of balance between static and dynamic organizational behaviors. The authors applied their model to traffic monitoring application. To our best knowledge, there is no any other real world application designed with MACODO.

### 4.2. Discussion

In MAS literature, there is large number of organizational models proposed by MAS researchers from all over the world to support the organizational aspects within MAS. Some of these models adopt the ACMAS viewpoint, others are concerned with the OCMAS viewpoint, and others adopt a hybrid approach combines both viewpoints. For the sake of this article size, it is not possible to explore all proposed organizational models in details, but interested readers can see [70], which is a handbook of research on MAS organizational models contains many of recent organizational model. Bellow, we provide our remarks about these models:
- This large number of organizational models indicates that concerning organizational aspects within MAS is currently a very interesting research area.
- It also emphasize that till now there is no a fit-to-all organizational models that can be used to design MAS-based systems in all application domains.
- Nearly, each of these models was dedicated to specific real world application domain and it is not applied to other applications.

- Some of these organizational models tackle with organization structure issues at design time (pure static), and others tackle them at will (pure dynamic).
- In some of them the organization abstraction is not explicit and the responsibility of dynamic reorganization is given to individual agents in addition to their functional responsibilities.
- Most of them considered the intra-organization and did not tackle with inter-organization reorganization. How to model the interaction among organizations?
- Few organizational models tackled both static and dynamic aspects of organizations and environments.
- In most of them the individual agent initiates to join a certain organization and this require that the agent has a reasonable knowledge about the services of each organization to select the appropriate one to join. We argue that letting the organization itself to select suitable agents to award it a role is the better approach because organization knowledge is more global than that of an individual agent.

Based on these remarks and limitations of most previously proposed MAS organizational models, we proposed a novel organizational model for engineering complex large-scale MAS-based applications. The new MAS organizational model was called NOSHAPE and conceptually presented in [76]. NOSHAPE is supposed to handle all the limitations of other related models. It exploits the overlapping relationships among higher order abstraction entities such as organizations of agents, worlds of organizations, and even universes of worlds within MAS to realize and utilize their captivating characteristics.

## 5. Conclusion and Future Work

MAS organization can be considered as a process to dynamically reorganize the system-to-be to adapt environment dynamic changes. Or, it can be considered as an entity facilitates the partitioning of the system-to-be. Organizations are a typical way to structure and manage interactions among agents. Establishing an organizational structure that specifies how agents in a system should work together helps the achievement of effective coordination in MAS. This chapter provided a comprehensive overview about MAS organization including its motivations, paradigms, models, and other related concepts such as self-organization and emergence. In MAS literature, we found very large number of organizational models proposed to support dynamic reorganization of the MAS, this large number of organizational models indicates that concerning organizational aspects within MAS is currently a very active and interesting research area and that till now there is no a fit-to-all MAS organizational model that can be used for engineering all possible application domains. This conclusion motivates us to propose a novel organizational model for engineering complex and highly distributed large-scale MAS such as modern industrial networks (i.e., SCADA [19]).

## References

[1] Shehory, O. Architectural properties of multi-agent systems. Technical Report CMU-RI-TR-98-28, The Robotics Institute, Carnegie Mellon University, Pittsburgh, Pennsylvania 15213, 1998.

[2] Jennings, N. R., & Wooldridge, M. Agent-Oriented Software Engineering. in Bradshaw, J. ed. Handbook of Agent Technology, AAAI/MIT Press, 2000.

[3] Gasser, L. (2001). Perspectives on Organizations in Multi-Agent Systems. In Multi-Agent Systems and Applications, Michael Luck et al (pp. 1–16). Berlin: Springer-Verlag. doi:10.1007/3-540-47745-4_1.

[4] Odell, J. J., Parunak, H. V. D., & Fleischer, M. (2003). The role of roles in designing effective agent organizations. In Software Engineering for Large-Scale Multi-Agent Systems (pp. 27–38). Springer Berlin Heidelberg. doi:10.1007/3-540-35828-5_2.

[5] Horling, B., & Lesser, V. (2004). A survey of multi-agent organizational paradigms. The Knowledge Engineering Review, 19(4), 281–316.doi:10.1017/ S0269888905000317.

[6] Ferber, J., Gutknecht, O., & Michel, F. (2004). From agents to organizations: an organizational view of multi-agent systems. In Agent-Oriented Software Engineering IV (pp. 214–230). Springer Berlin Heidelberg. doi:10.1007/978-3-540-24620-6_15.

[7] Van Den Broek, E. L., Jonker, C. M., Sharpanskykh, A., & Treur, J. (2006). Formal modeling and analysis of organizations. In Coordination, Organizations, Institutions, and Norms in Multi-Agent Systems (pp. 18-34). Springer Berlin Heidelberg.

[8] Hübner, J. F., Vercouter, L., & Boissier, O. (2009). Instrumenting multi-agent organisations with artifacts to support reputation processes. In Coordination, Organizations, Institutions and Norms in Agent Systems IV (pp. 96–110). Springer Berlin Heidelberg. doi:10.1007/978-3-642-00443-8_7.

[9] Ferber, J., Michel, F., & Báez, J. (2005). AGRE: Integrating environments with organizations. In Environments for multi-agent systems (pp. 48-56). Springer Berlin Heidelberg.

[10] Dignum, V. (2009). The role of organization in agent systems. Handbook of Research on Multi-Agent Systems: Semantics and Dynamics of Organizational Models, 1-16.

[11] Weyns, D., Haesevoets, R., & Helleboogh, A. (2010). The MACODO organization model for context-driven dynamic agent organizations. ACM Transactions on Autonomous and Adaptive Systems (TAAS), 5(4), 16.

[12] Burns, T., & Stalker, G. (1961). *The Management of Innovation*, Tavistock, London.

[13] Picard, G., Hübner, J. F., Boissier, O., & Gleizes, M. P. (2009, June). Reorganisation and self-organisation in multi-agent systems. In 1st International Workshop on Organizational Modeling, ORGMOD (pp. 66-80).

[14] Upton, J., Janeka, I., & Ferraro, N. (2014). The whole is more than the sum of its parts: aristotle, metaphysical. Journal of Craniofacial Surgery, 25(1), 59-63.

[15] A. Drogoul, B. Corbara, and S. Lalande. MANTA: New experimental results on the emergence of (artificial) ant societies. In Nigel Gilbert and Rosaria Conte, editors, Artificial Societies: the Computer Simulation of Social Life, pages 119–221. UCL Press, London, 1995.

[16] Rajkumar, R. R., Lee, I., Sha, L., & Stankovic, J. (2010, June). Cyber-physical systems: the next computing revolution. In Proceedings of the 47th Design Automation Conference (pp. 731-736). ACM.

[17] Mattern, F., & Floerkemeier, C. (2010). From the Internet of Computers to the Internet of Things. In From active data management to event-based systems and more (pp. 242-259). Springer Berlin Heidelberg.

[18] James Momoh (2012), Smart Grid: Fundamentals of Design and Analysis, Wiley-IEEE Press; 1 edition (March 20, 2012).

[19] Abbas, H. A. (2014). Future SCADA challenges and the promising solution: the agent–based SCADA. *International Journal of Critical Infrastructures*, *10*(3), 307-333.

[20] Saha, D., & Mukherjee, A. (2003). Pervasive computing: a paradigm for the 21st century. *Computer*, *36*(3), 25-31.

[21] Friedewald, M., & Raabe, O. (2011). Ubiquitous computing: An overview of technology impacts. *Telematics and Informatics*, *28*(2), 55-65.

[22] Jennings NR (2001) An agent-based approach for building complex software systems. Communications of the ACM 44 (4):35–41

[23] Ferber J (1999) Multi-Agent Systems: An Introduction to Distributed Artificial Intelligence. Addison-Wesley, Harlow, England.

[24] Wooldridge, M.: An Introduction to Multi-Agent Systems. Wiley, New York (2002).

[25] Posey, Rollin B. (March 1961). "Modern Organization Theory edited by Mason Haire". Administrative Science Quarterly 5 (4): 609–611.

[26] Hertz, D. and R. Livingston. (1950). Contemporary Organizational theory: A review of current concepts and methods. Human Relations, 3 (4), 373-394.

[27] Odell, J. J., Parunak, H. V. D., & Fleischer, M. (2003). The role of roles in designing effective agent organizations. In *Software Engineering for Large-Scale Multi-Agent Systems* (pp. 27-38). Springer Berlin Heidelberg.

[28] Van Den Broek, E. L., Jonker, C. M., Sharpanskykh, A., & Treur, J. (2006). Formal modeling and analysis of organizations. In *Coordination, Organizations, Institutions, and Norms in Multi-Agent Systems* (pp. 18-34). Springer Berlin Heidelberg.

[29] Miles, R. E., Snow, C. S., Mathews, J. A., Miles, G., & Coleman, H. J. (1997). Organizing in the knowledge age: Anticipating the cellular form. *The Academy of Management Executive*, *11*(4), 7-20.

[30] Lesser, V. R., & Corkill, D. D. (1981). Functionally accurate, cooperative distributed systems. Systems, Man and Cybernetics, IEEE Transactions on,11(1), 81-96.

[31] Corkill, D. D. (1980). *An organizational approach to planning in distributed problem-solving systems*. Technical Report 80-13, Department of Computer and Information Science, University of Massachusetts, Amherst, Massachusetts 01003.

[32] Jay Galbraith. *Designing Complex Organizations*. Addison-Wesley, 1973.

[33] Fox, M. S. (1981). An organizational view of distributed systems. *Systems, Man and Cybernetics, IEEE Transactions on*, *11*(1), 70-80.

[34] Giovanna Di Marzo Serugendo et al (2011), "Self-organizing Software, From Natural to Artificial Adaptation", Springer.

[35] M. Carley and L. Gasser. Computational organization theory. In G. Weiss, editor, Multiagent Systems: A Modern Approach to Distributed Arti_cial Intelligence, pages 299.330. MIT Press, 1999.

[36] Corkill, D. D., & Lander, S. E. (1998). Diversity in agent organizations. *Object Magazine*, *8*(4), 41-47.

[37] Jay R. Galbraith. *Organization Design*. Addison-Wesley, 1977.

[38] Giovanna di Marzoserugendo et al, "Self-organization in multi-agent systems", The Knowledge Engineering Review, Vol. 20:2, 165–189., 2005, Cambridge University Press.

[39] K. S. Barber and C. E. Martin, 'Dynamic reorganization of decisionmaking groups', in Proceedings of the 5th Autonomous Agents, (2001).

[40] Dignum, V., Dignum, F., & Sonenberg, L. (2004, September). Towards dynamic reorganization of agent societies. In *Proceedings of Workshop on Coordination in Emergent Agent Societies at ECAI* (pp. 22-27).

[41] Bernon, C., Camps, V., Gleizes, M. P., & Picard, G. (2005). Engineering adaptive multi-agent systems: The adelfe methodology. Agent-oriented methodologies, 172-202.

[42] Guessoum, Z., Briot, J. P., Marin, O., Hamel, A., & Sens, P. (2003). Dynamic and adaptive replication for large-scale reliable multi-agent systems. InSoftware engineering for large-scale multi-agent systems (pp. 182-198). Springer Berlin Heidelberg.

[43] Berns, A., & Ghosh, S. (2009, September). Dissecting self-* properties. In Self-Adaptive and Self-Organizing Systems, 2009. SASO'09. Third IEEE International Conference on (pp. 10-19). IEEE.

[44] Sichman, J. S., Dignum, V., & Castelfranchi, C. (2005). Agents' organizations: a concise overview. Journal of the Brazilian Computer Society, 11(1), 3-8.

[45] Serugendo, G. D. M., Gleizes, M. P., & Karageorgos, A. (2006). Self-Organisation and Emergence in MAS: An Overview. Informatica (Slovenia),30(1), 45-54.

[46] P. Glansdorff and I. Prigogine. Thermodynamic study of Structure, Stability and Fluctuations. Wiley, 1971.

[47] Mano, J. P., Bourjot, C., Lopardo, G., & Glize, P. (2006). Bio-inspired mechanisms for artificial self-organised systems. Informatica, 30(1), 55-62.

[48] Giovanna (2011) Di Marzo Serugendo et al, "Self-organizing Software, From Natural to Artificial Adaptation", Springer, 2011.

[49] Zambonelli, F., Gleizes, M. P., Mamei, M., & Tolksdorf, R. (2004, May). Spray computers: frontiers of self-organization. In Autonomic Computing, 2004. Proceedings. International Conference on (pp. 268-269). IEEE.

[50] Karuna, H., Valckenaers, P., Saint-Germain, B., Verstraete, P., Zamfirescu, C. B., & Van Brussel, H. (2005). Emergent forecasting using a stigmergy approach in manufacturing coordination and control. In Engineering Self-Organising Systems (pp. 210-226). Springer Berlin Heidelberg.

[51] Weyns, D., Schelfthout, K., Holvoet, T., & Glorieux, O. (2004). Role based model for adaptive Agents. In *Fourth Symposium on Adaptive Agents and Multiagent Systems at the AISB'04 Convention*.

[52] M.P. Gleizes, V. Camps, and P. Glize. A theory of emergent computation based on cooperative self-organisation for adaptive artificial systems. Fourth European Congress of Systems Science. Valencia, 1999.

[53] De Wolf, T., & Holvoet, T. (2004). Emergence and self-organisation: a statement of similarities and differences. *Engineering Self-Organising Systems*, *3464*, 1-15.

[54] Akgün, A. E., Keskin, H., & Byrne, J. C. (2014). Complex adaptive systems theory and firm product innovativeness. Journal of Engineering and Technology Management, 31, 21-42.

[55] Streng, W. (2005). Reductionism versus Holism–Contrasting Approaches.Consilience. Interdisciplinary Communications, 2006, 11-14.

[56] Ashby, W.R.: Principles of self-organizing dynamic systems. Journal of General Psychology 37 (1947) 125–128.

[57] Giorgini, P., & Henderson-Sellers, B. (2005). Agent-oriented methodologies: an introduction. *Agent-oriented Methodologies*, 1-19.

[58] Ferber, J. and Gutknecht, O., Aalaadin: a meta-model for the analysis and design of organizations in multi-agent systems. in Third International Conference on Multi-Agent Systems, (Paris, 1998), IEEE, 128-135.

[59] Wooldridge, M., Jennings, N. R., & Kinny, D. (2000). The Gaia methodology for agent-oriented analysis and design. Autonomous Agents and Multi-Agent Systems, 3(3), 285-312.

[60] Ferber, J., Michel, F., & Báez, J. (2005). AGRE: Integrating environments with organizations. In Environments for multi-agent systems (pp. 48-56). Springer Berlin Heidelberg.

[61] Weyns, D., Haesevoets, R., & Helleboogh, A. (2010). The MACODO organization model for context-driven dynamic agent organizations. ACM Transactions on Autonomous and Adaptive Systems (TAAS), 5(4), 16.

[62] Spivey, J. M. (1989). The Z notation (Vol. 1992). New York: Prentice Hall.

[63] Coutinho, L. R., Sichman, J. S., & Boissier, O. (2005, October). Modeling organization in mas: A comparison of models. In First Workshop on Software Engineering for Agent-oriented Systems (pp. 1-10).

[64] Argente, E., Palanca, J., Aranda, G., Julian, V., Botti, V., Garcia-Fornes, A., & Espinosa, A. (2007). Supporting agent organizations. In Multi-Agent Systems and Applications V (pp. 236-245). Springer Berlin Heidelberg.

[65] Dignum, V., Dignum, F.: A Landscape of Agent Systems for the Real World. Tech. Report Utrecht University (2007)

[66] Gutknecht, O., & Ferber, J. (1998). A model for social structures in multi-agent systems (Vol. 98040). Technical Report RR LIRMM.

[67] Hannoun, M., Boissier, O., Sichman, J. S., & Sayettat, C. (2000). MOISE: An organizational model for multi-agent systems. In Advances in Artificial Intelligence (pp. 156-165). Springer Berlin Heidelberg.

[68] Hübner, J. F., Sichman, J. S., & Boissier, O. (2002, July). Moise+: towards a structural, functional, and deontic model for mas organization. In Proceedings of the first international joint conference on Autonomous agents and multiagent systems: part 1 (pp. 501-502). ACM.

[69] Hübner, J. F., Sichman, J. S., & Boissier, O. (2004). Using the MOISE$^+$ for a Cooperative Framework of MAS Reorganisation. In Advances in artificial intelligence–SBIA 2004 (pp. 506-515). Springer Berlin Heidelberg.

[70] Dignum, V. (Ed.). (2009). Handbook of Research on Multi-Agent Systems: Semantics and Dynamics of Organizational Models: Semantics and Dynamics of Organizational Models. IGI Global.

[71] Jennings, N. R. (1999). Agent-oriented software engineering. In Multiple Approaches to Intelligent Systems (pp. 4-10). Springer Berlin Heidelberg.

[72] Duan, Junhua, Yi-an Zhu, and Shujuan Huang. "Stigmergy agent and swarm-intelligence-based multi-agent system." Intelligent Control and Automation (WCICA), 2012 10th World Congress on. IEEE, 2012.

[73] Smith, R. G. (1980). The contract net protocol: High-level communication and control in a distributed problem solver. IEEE Transactions on computers, (12), 1104-1113.

[74] S.A. DeLoach. Methodologies and Software Engineering for Agent Systems. The Agent-Oriented Software Engineering Handbook Series : Multiagent Systems, Artificial Societies,and Simulated Organizations, volume 11, chapter The MaSE Methodology. Kluwer Academic Publishing (available via Springer), 2004.

[75] Juan Pavón and Jorge J. Gómez-Sanz. Agent oriented software engineering with ingenias. In Vladimír Mařík, Jörg P. Müller, and Michal Pechoucek, editors, CEEMAS, volume 2691 of Lecture Notes in Computer Science, pages 394–403. Springer, 2003.

[76] Abbas, H. A. (2014). Exploiting the Overlapping of Higher Order Entities within Multi-Agent Systems. International Journal of Agent Technologies and Systems (IJATS), 6(3), 32-57. doi:10.4018/ijats.2014070102.